# A Shallow U-Net Architecture for Reliably Predicting Blood Pressure (BP) from Photoplethysmogram (PPG) and Electrocardiogram (ECG) Signals

Sakib Mahmud[1], Nabil Ibtehaz[1], Amith Khandakar[1], Anas Tahir[1], Tawsifur Rahman[1], Khandaker Reajul Islam[1], Md Shafayet Hossain[2], M. Sohel Rahman[3], Mohammad Tariqul Islam[2], Muhammad E. H. Chowdhury*[1]

[1]Department of Electrical Engineering, Qatar University, Doha 2713, Qatar

[2]Department of Electrical, Electronics and Systems Engineering, Universiti Kebangsaan Malaysia, Bangi, Selangor 43600, Malaysia

[3]Department of CSE, BUET, ECE Building, West Palashi, Dhaka-1205, Bangladesh

Corresponding author: Muhammad E. H. Chowdhury, (email: mchowdhury@qu.edu.qa)

## Abstract
Cardiovascular diseases are the most common causes of death around the world. To detect and treat heart-related diseases, continuous Blood Pressure (BP) monitoring along with many other parameters are required. Several invasive and non-invasive methods have been developed for this purpose. Most existing methods used in the hospitals for continuous monitoring of BP are invasive. On the contrary, cuff-based BP monitoring methods, which can predict Systolic Blood Pressure (SBP) and Diastolic Blood Pressure (DBP), cannot be used for continuous monitoring. Several studies attempted to predict BP from non-invasively collectible signals such as Photoplethysmogram (PPG) and Electrocardiogram (ECG), which can be used for continuous monitoring. In this study, we explored the applicability of autoencoders in predicting BP from PPG and ECG signals. The investigation was carried out on 12,000 instances of 942 patients of the MIMIC-II dataset and it was found that a very shallow, one-dimensional autoencoder can extract the relevant features to predict the SBP and DBP with the state-of-the-art performance on a very large dataset. Independent test set from a portion of the MIMIC-II dataset provides an **MAE** of **2.333** and **0.713** for SBP and DBP, respectively. On an external dataset of forty subjects, the model trained on the MIMIC-II dataset, provides an **MAE** of **2.728** and **1.166** for SBP and DBP, respectively. For both the cases, the results met British Hypertension Society (BHS) Grade A and surpassed the studies from the current literature.

## I. Introduction
Despite tremendous advancements in the healthcare sector, cardiovascular diseases (CVDs) still secure the top positions last year in the list of leading causes of death globally. The most fatal CVD is the "Ischaemic Heart Disease" which is termed by the World Health Organization (WHO) as the "World's Biggest Killer" as it accounted for 16% of the total deaths from 2000 to 2019 [1]. The second, third, and fourth positions are secured by "Stroke", "Chronic Pulmonary Diseases" and "Lower Respiratory Infections", respectively which are also, directly and indirectly, related to CVDs [2-4]. Hypertension or High Blood Pressure (BP) is one of the leading causes of CVDs: almost 54% of strokes and 47% of coronary heart diseases, worldwide, can be attributed to high BP [5]. In the USA alone, there are around 67 million people (almost one-third of the population) suffering from various hypertension problems while the irony is, according to this statistic [6], more than half of them are reluctant to mitigate their condition. The main reason behind this kind of reluctance seen among high BP patients is the dormant nature of hypertension which eventually leads to untimely death. For this reason, it is commonly termed the "Silent Killer" [7]. Due to the silent nature of hypertension, it is crucial to continuously monitor the BP of the patients. Due to a shortage of expert physicians compared to the huge number of patients; automated BP monitoring methods seem to be a viable alternative in this regard.

Several studies attempted to tackle this problem following various methodologies and a handful of them (e.g., [8]) were adopted by the healthcare centers to measure blood pressure continuously. But most of the robust methods for blood pressure monitoring are either intermittent or invasive if continuous. Two commonly used BP recording methods are the cuff-based oscillometric technique and arterial blood pressure (ABP) reading from the radial artery through cannulation. Both of these methods are reliable; the former is non-invasive albeit intermittent whereas the latter is continuous but invasive [9]. A few calibration-based techniques have been developed to detect BP non-invasively, such as the Photoplethysmography (PPG) based Finger-Clamp method [10] and the Applanation Tonometry method [11]. Calibration is required for both of these non-invasive techniques since they do not readily provide the correct BP values or ABP signals. This type of calibration or mapping can also be useful in measuring BP readings [12] or ABP waveforms [13] from PPG if they are recorded simultaneously. Recently, a few studies have been reported where various Machine Learning (ML) and statistical techniques are used to predict BP from non-invasively collected PPG signals. Some studies used traditional ML (regression) models, such as, Support Vector Regressor (SVR) [14], Adaptive Boosting (AdaBoost) [12], Random Forest [15], Gradient Boosting (GradBoost) [16], Gaussian Process Regression (GPR) [18], Artificial Neural Network (ANN) [20-23], Recurrent Neural Network (RNN) based Long Short-Term Memory (LSTM) [24, 25], etc. on small or medium-sized datasets to predict BP from PPG alone or a combination of PPG and electrocardiogram (ECG).

In recent years, Convolution Neural Network (CNN) based Deep Learning (DL) techniques have been utilized to solve complex problems on large datasets in 1D (e.g., Signals), 2D (e.g., Images, and even in 3D (e.g., Videos) settings. However, there are not many deep learning-based approaches in the literature for BP estimation. Slapničar et al. [26] predicted BP from PPG, its derivatives (1D signals) and their respective spectrograms (2D signals) using a hybrid pipeline containing both 1D and 2D CNNs termed as "Spectro-Temporal ResNets". Recently, Athaya et al. [27] used modified U-Net [28] architecture for PPG to ABP signal to signal translation. On the other hand, Ibtehaz et al. [13] in their work used two CNN networks in sequence (U-Net and MultiResUNet [29]) for PPG to ABP signal translation but could not reach Grade A [58] for Systolic Blood Pressure (SBP) prediction. So, based on the current literature, there is still scope for significant improvement in BP predictions using deep learning models.

U-Net is an encoder-decoder-based deep CNN architecture that was originally used for image (2D) segmentation. Many studies used U-Net to perform tasks, such as Biomedical Image Segmentation [30], Shape Regeneration [31], Road Shape Extraction from Satellite Maps [32], etc. A good number of studies tried to design special-purpose variations of U-Net such as U-Net++ [30], nnu-Net [33], Ternausnet [34], Wave-U-Net [35], etc. for various applications, mostly in 2D settings. There is also a 3D version of U-Net [36] designed for tackling three-dimensional problems. In the 1D domain, apart from the PPG to ABP signal translation discussed earlier, there have also been works in speech enhancement [37], echo cancellation [38], heartbeat detection, etc. Thus, the U-Net architecture has been modified in various ways for solving different types of problems and in a few cases, a shallow U-Net performed better than a deeper version. For example, Wu et al. [39] utilized a shallow three-layer version of U-Net used for shadow detection as part of a scene understanding task. On the other hand, to the best of our knowledge, the U-Net architecture has rarely been used just for feature extraction while acting as an autoencoder. Esmaelpoor et al. [40] in their study tried to extract features from PPG signals using a generic CNN with two convolutional layers, then used those features on LSTM models to predict BP. Features were extracted separately from PPG and ECG and both were put into LSTM networks to separately predict SBP and Diastolic Blood Pressure (DBP). In this study, we followed a similar approach but for feature extraction, we utilized the encoder portion of the U-Net. A densely connected Multi-Layer Perceptron (MLP) layer was added to the end of the encoder for extracting the network learned features. This lightweight version of the U-Net can easily be applied to devices in a (computing and memory) resource-constrained setting. Thus, the novelty of this work lies not only in the feature extraction pipeline but also in using the shallowest version of U-Net on a large dataset for extracting features optimizing the BP prediction process. To the best of our knowledge, our extracted latent features from the shallowest U-Net have outperformed all the state-of-the-art techniques hitherto found in the literature.

## II. Materials and Methods
### A. Datasets
In this study, two different datasets have been used, which are briefly described below.
#### 1) MIMIC-II Dataset from the UCI Repository:
The Cuff-Less Blood Pressure Estimation Dataset [14] from the UCI Machine Learning Repository [41], termed as the "UCI Dataset", has been used in this study. The UCI Dataset is a filtered and processed version of the Multi-Parameter Intelligent Monitoring in Intensive Care II (MIMIC-II) Waveform database [42,43]. The MIMIC-II Waveform database contains records of continuous high-resolution physiologic waveforms and minute-by-minute numeric trends of physiologic measurements, such as ABP, PPG, Cerebral Perfusion Pressure (CPP), Central Venous Pressure (CVP), Pulmonary Arterial Pressure (PAP), so on and so forth. The UCI Dataset contains 12000 instances of simultaneous PPG, ABP, and ECG data of 942 patients extracted from the MIMIC-II Waveform database with a sampling rate of 125 Hz. The 12000 instances of the UCI Dataset were uniformly divided into four parts, each part containing 3000 instances, and the data is available in MATLAB file format (". mat"). Even though the MIMIC-II database has data from a large number of patients, only 942 patients had all three PPG, ECG, and ABP signals simultaneously, which is required for BP prediction in the proposed model. UCI Dataset was created with only the MIMIC-II records where all three of PPG, ABP, and ECG data were present. While creating the UCI Dataset, Kachuee et al. [14] performed some signal processing tasks, such as smoothing all signals using a simple averaging filter, removing signals with unacceptable human Blood Pressure (BP) and Heart Rate (HR) values, getting rid of signals with severe discontinuities and auto-correlating PPG signals for checking the similarity between successive pulses. Therefore, these steps were not repeated in this study.

#### 2) BCG Dataset:
The external validation dataset used in this work has been collected and shared recently by Carlson et al. [44] (referred to as "BCG Dataset" in this paper). Several heart-driven signals, such as Ballistocardiogram (BCG), ECG, PPG, and ABP waveforms are available in the dataset. Note that, BCG waveforms of this dataset are not of any interest for this study. Data were collected from 40 subjects (17 males and 23 females) with a sampling rate of 1000 Hz. The signals were digitized by the NI-9220 [45] device, which was used to gather signals collected by various data acquisition devices. The ABP signals in this dataset were non-invasively collected from the reconstructed brachial artery pressure (reBAP) signals, which were collected using Finometer Pro [46] from Finapres Medical Systems. The ABP signals were represented in terms of volts following a normalizing scale of 100 mmHg/volt. The BCG Dataset is also available in the ".mat" file format. An overview of both datasets is provided in Table 1.

**Table 1.** Overview of the Datasets (after pre-processing)

| Datasets | BP Parameters | Minimum | Maximum | Mean | Standard Deviation |
|---|---|---|---|---|---|
| UCI Dataset | SBP | 80.026 | 189.984 | 132.609 | 21.703 |
|  | DBP | 50.000 | 119.927 | 63.705 | 9.978 |
|  | MAP | 57.941 | 149.062 | 87.228 | 12.737 |
| BCG Dataset | SBP | 80.313 | 186.641 | 124.535 | 15.237 |
|  | DBP | 43.899 | 96.829 | 65.011 | 9.180 |
|  | MAP | 62.975 | 128.391 | 86.878 | 10.046 |

As seen from Table 1, the UCI dataset, even though larger, is more deviated, especially for SBP. On the other hand, DBP and mean arterial pressure (MAP) of the signals in the UCI dataset vary within a much wider range than the ones in the BCG dataset. The sampling rate of the signals in the BCG dataset was resampled at 125 Hz from 1000 Hz to maintain harmony with the UCI dataset signals. So, the duration of a sample signal was about (1024/125) ≈ 8.192 seconds. It means that the total duration of the data collected from the UCI dataset was about **456 hours** and this is around **4.26 hours** for the BCG dataset.

### B. Data Pre-processing

At first, the signal was segmented to 1024 samples from the UCI dataset while preserving the original sampling rate of 125 Hz. Signals from the UCI dataset suffer from severe baseline drift in many instances. So, baseline wandering was removed before normalizing the signals. After fixing the baseline drifts and properly normalizing the signals, the first two derivatives of PPG were derived and stored along with their corresponding PPG signals to be used as predictors alongside PPG and ECG. Before compiling the whole dataset, highly distorted signals were removed. Signal pre-processing was performed in MATLAB (version R2020a). The whole data pre-processing procedure is shown in Figure 1. The BCG dataset was also pre-processed similarly. But before pre-processing, their sampling frequency was down-sampled from 1000Hz to 125 Hz to ensure consistency with the UCI dataset. The ABP signals in the BCG dataset were denormalized by multiplying with a factor of 100 since they were normalized and stored by maintaining a scale of 100 mmHg/volt.

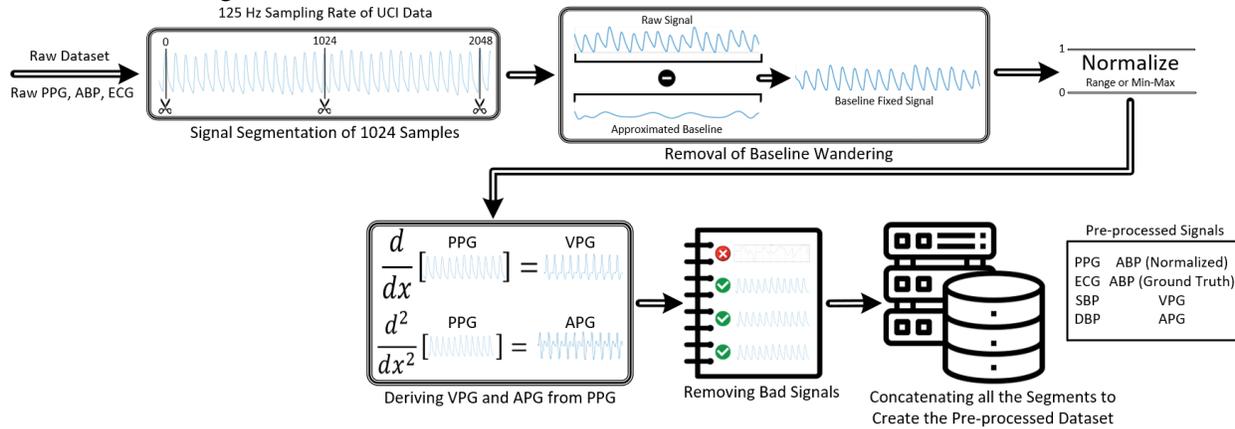

**Figure 1:** Flowchart representing the data pre-processing pipeline for the UCI Dataset. The pipeline for the BCG Dataset is almost identical except the ABP signals were denormalized first by multiplying with the normalizing factor of 100.

**Baseline Drift Correction:** Baseline drift correction was done using the built-in functions of MATLAB ('movmin' [47], 'polyfit', and 'polyval' [48]). At first 'movmin' or moving minimum function was used to find an array of estimated minimum points acting as a baseline approximation for the waveform. Afterward, the 'polyfit' function was used to fit a higher-order polynomial along with the estimated points and 'polyval' [46] was used to formulate the polynomial based on the 'polyfit' result, which is the estimated baseline. Then the baseline is deducted from the raw signal to get the baseline drift corrected signal.

**Normalization:** PPG and ECG were range normalized between 0 and 1 per signal while ABP waveforms were min-max normalized globally, in terms of the minimum and maximum of the ABP waveforms across the whole dataset. ABP signals were not range normalized between 0 to 1 to retain their relative amplitude feature (i.e., BP levels) which was found to be helpful during BP prediction.

**Derivatives of PPG:** According to literature, the first and second derivatives of PPG also provide valuable information or features while predicting BP. They are called in various names such as PPG', PPG'' or Velocity of PPG (VPG), Acceleration of PPG (APG), or FDPPG (First Derivative of PPG), SDPPG (Second Derivative of PPG) [47-49] (Figure 2). To find the VPG and APG from PPG, MATLAB's 'diff' function has been used. But a finite "Step Size" [50] of the "diff" function induces distortions in the derived signals which keep increasing for higher-order derivatives. To remove these high-frequency distortions, the signals need to be filtered in each stage, which was done using MATLAB's "designfilt" function [51,52]. The cutoff frequencies for the bandpass filter were set carefully to pass through important frequency components related to PPG derivatives while attenuating low and high-frequency distortions. But applying a filter on the signals creates some delay which deteriorates along with the derivative order (APG > VPG). MATLAB's built-in function 'grpdelay' [56] was used to find the average filter delay. Then the signals were moved to the left by the amount of their respective delay. Adjustment of the length of original PPG signals was done to ensure the length of VPG and APG signals after the delay to maintain the length of the signals to be 1024.

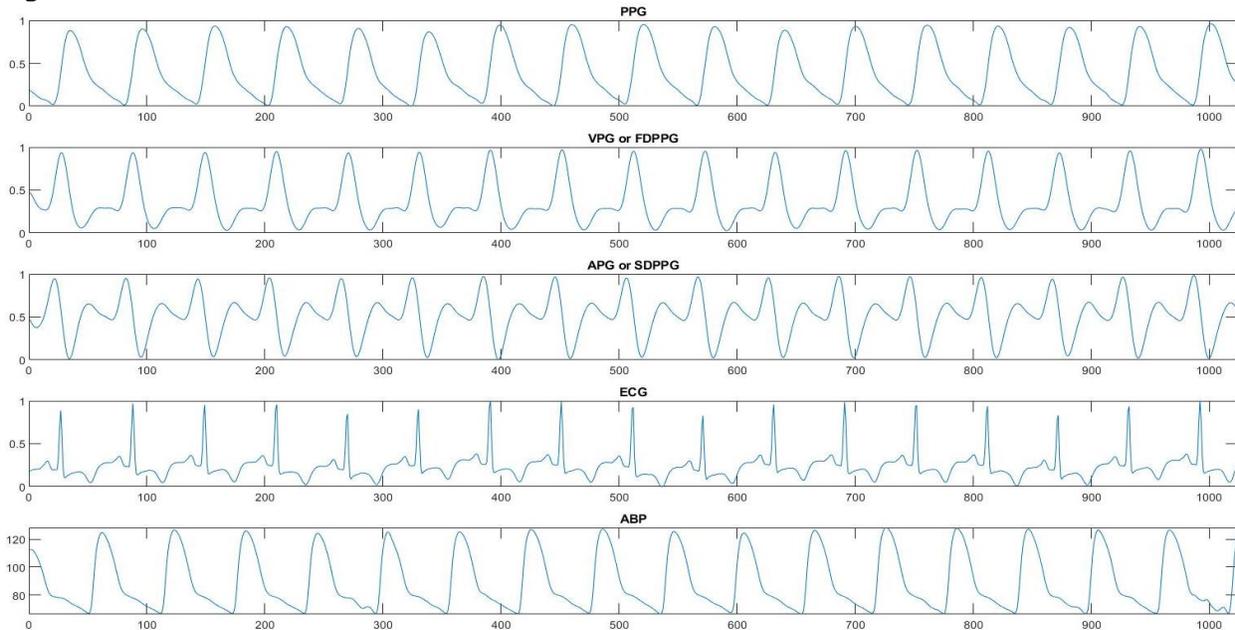

**Figure 2:** Snapshot of five signals (PPG, VPG, APG, ECG, and ABP) from a segment of the UCI dataset after pre-processing.

**Removing Bad Signals:** The signal samples extracted from the UCI dataset contains many highly distorted signals that can potentially affect the performance of the deep learning model significantly as the network tries to learn from them. Hence, the following types of samples were removed from the dataset - ABP Signals with extreme SBP and DBP values, Blank samples, and signals which exceed a certain distortion threshold. In particular, ABP signals with SBP values smaller than 80 and greater than 190, DBP values greater than 120 and smaller than 50, and ABP signals which had a BP range (SBP – DBP) less than 20 or more than 120 were removed since it was observed that apart from some extreme cases, highly distorted signals normally had such a BP range. Under this scheme, around 2% of signals got removed from the datasets. After performing some signal processing and taking the derivatives, a few samples became blank due to being extremely distorted; these were also removed. There are levels of distortions for various samples and a sample remains acceptable up to a certain level of distortion. As shown in Supplementary Figure 1, for ABP and PPG signals, the distorted samples had two main traits viz. highly non-uniform peaks either in terms of distance or height, and double peaks. Standard Deviation (STD) of the peak-to-peak distances and peak prominences (relative height) were observed to detect this anomaly and signals were sorted out based on a threshold of deviation. This threshold was set after performing trial and error by manually observing more than a thousand samples.

Histograms of ABP and SBP in Figure 3(a, c) can be compared for the signal distribution before and after the signal pre-processing. The box plots in Figure 3 show that after removing the low-quality signals, the number of outliers decreased, and a greater portion of signals entered into the interquartile range. Removing these outliers might improve the performance of the network. The median and standard deviation have changed marginally as the signal distribution is spreading more. Around 25% of both train and test signals were removed through this "bad signal removal" scheme. It is mentionable that most other researchers also worked on ABP signals of a certain BP range alongside putting on other constraints to boost the network performance [13,22,24,25,26]. Even though a good number of segments were removed, due to using more than one channel and considering the whole UCI version of the MIMIC-II dataset, a comparatively larger number of segments were available for training, validation, and testing.

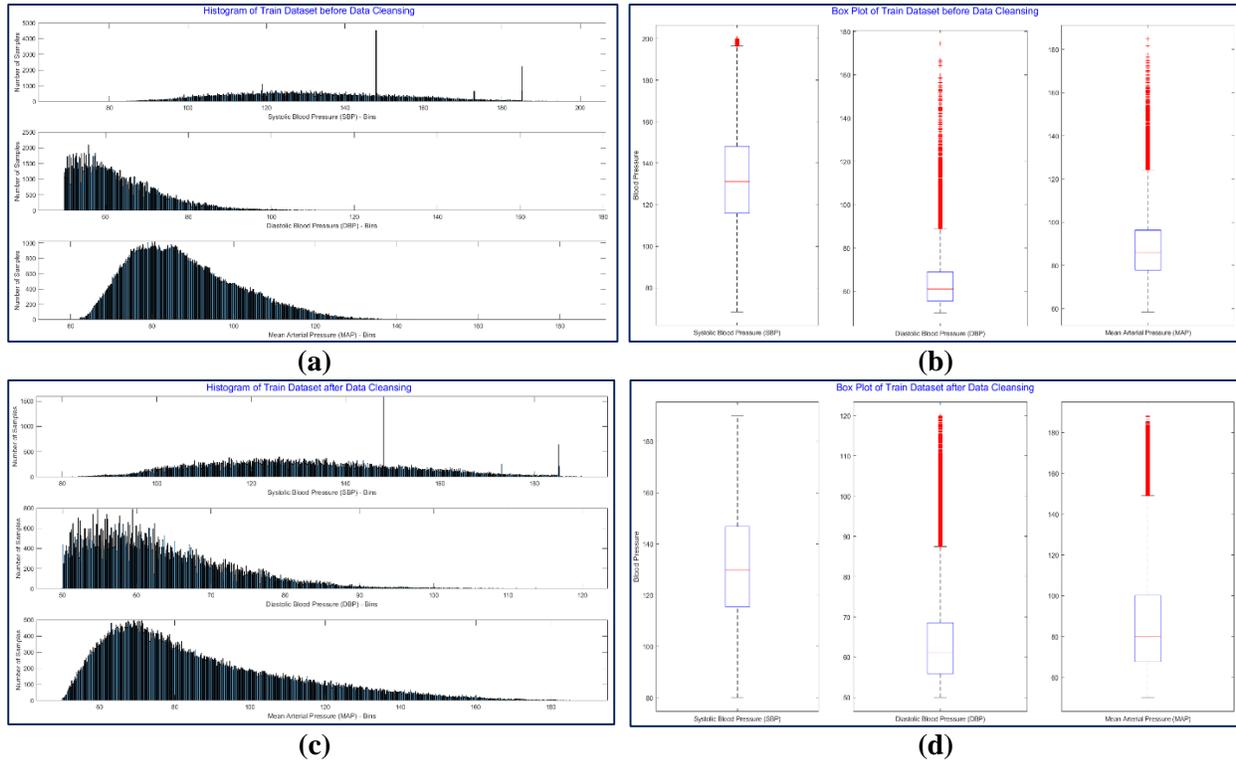

**Figure 3:** Histogram and Box Plot(s) of SBP, DBP, and MAP of Signals in the Training Set before and after Data Cleansing

## C. Rationale behind this Study

The rationale behind this study was to extract an effective set of features from a very large dataset containing PPG, ECG, and ABP signals which can be used to reliably predict BP. While studies mostly use PPG and ECG signals for BP prediction or extracting features directly, we propose to use an approach inspired by the power of autoencoders to extract the latent features automatically and check for the performance. In traditional autoencoders, usually, the input is given to the network to reconstruct it through a latent space compact representation. This enforces the model to learn the distinctive attributes of the input and thus has shown great success in feature extraction. Therefore, an obvious idea would be to train an autoencoder using PPG and ECG signals both as inputs and outputs. This will provide us with a latent space aware of the diverse patterns of the PPG and ECG signals, which should be able to predict BP correctly. However, rather than optimizing the feature space to reconstruct PPG and/ECG, if we can optimize the Unet to predict ABP in the output, we believe the feature space will be more optimized. Therefore, we had trained the autoencoder with PPG and ECG signals as input and ABP waveform as the output anticipating that the network will inherently learn to map the ECG, PPG signals to the ABP waveforms. Consequently, through this process, the network is expected to map the various patterns of the PPG and ECG signals to the corresponding patterns in the ABP signal. Therefore, we hypothesize that by applying the above-mentioned training mechanism of the autoencoder network, we can extract features from the PPG and ECG signals responsible for the changes in ABP. As a result, a regressor model, trained with these features, will likely be able to predict blood pressure better.

### D. Pipeline for BP Prediction
The BP prediction pipeline consists of mainly two sections viz. the U-Net based autoencoder for feature extraction from the raw signals, and the Machine Learning based regressor to perform regression on the extracted features for BP prediction. The complete pipeline is shown in Figure 4.

**Feature Extractor:**
The U-Net-based autoencoder is used for extracting a feature map from the raw input data. The dimensionality of the feature map may vary depending on the network setup (discussed elaborately in the experiments section). The training parameters for the UNet based feature extractor (autoencoder) are: Batch Size = 64, Number of epochs = 100, Patience (Stopping Criterion) = 15, Mean Squared Error (MSE) as the loss function, Adam as the optimizer and Mean Absolute Error (MAE) as the metric being monitored. Batch size, number of training epochs and the patience value were varied few times initially to determine their optimum values.

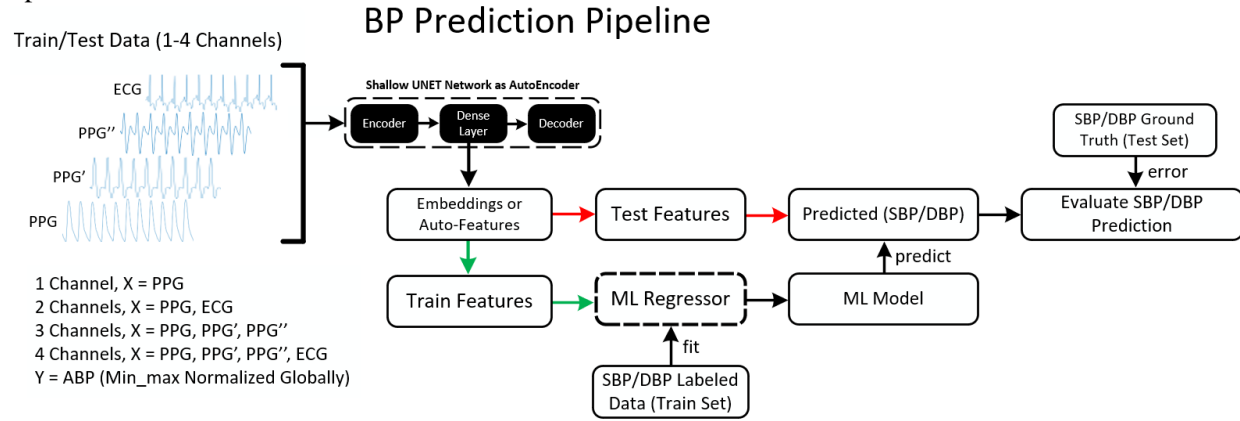

**Figure 4:** Complete Pipeline for BP Prediction using a Two-Stage Machine Learning Approach (Dash Means an Iterative Process)

**Regressor:**
Extracted features were regressed using traditional Machine Learning (ML) techniques, such as K-Nearest Neighbor (KNN), Support Vector Machines (SVM), Stochastic Gradient Descent (SGD), various Ensemble techniques (e.g., Adaptive Boosting, Gradient Boosting, Extreme Gradient Boosting (XgBoosting), and Random Forest), and Artificial Neural Network (ANN) based Multi-Layer Perceptron (MLP). For all these ML algorithms, various parameters were tweaked and tuned to get the optimum outcome. As shown in Supplementary Table 2, for MLP, Adam was chosen as the solver, ReLU as the activation function, Invscaling as the learner, alpha = 0.0001, batch size = auto, max iteration = 500 and hidden layer size = 100.

### III. Experiments
The primary aim of the experiments was to find the best performing U-Net architecture which can be used as an autoencoder for optimum feature extraction. Later, the same pipeline can be used to evaluate external datasets. So, mainly two types of experiments were performed in this study as discussed below.

### A. Experiment 1 (Train and Test on UCI Dataset):
The UCI dataset (12,000 instances) was originally divided into four equal parts. The first three parts of the UCI dataset were combined to make the train set (75% of the dataset – 9000 instances) and the fourth part was taken as an independent test set (25% of the dataset – 3000 Instances). During training, randomly selected 20% of the training set was used for validation. Four combinations of the four input signals viz. PPG, ECG, VPG, and APG were used in this experiment, while the target signal is ABP. Total predictor signal segments used for the four-channel approach (PPG, VPG, APG, and ECG as the four predictor signals) were 147,116 while the test set size was 53,043, as shown in Table 2.

Table 2. Description of Training and Testing Sets for Experiment 1

| No. of Channels | Channels | Target | Total Samples in the Train Set | Total Samples in the Test Set |
|---|---|---|---|---|
| 1 | PPG | ABP | 147,116 | 53,043 |
| 2 | PPG, ECG | ABP | 147,116 | 53,043 |
| 3 | PPG, VPG, APG | ABP | 147,116 | 53,043 |
| 4 | PPG, VPG, APG, ECG | ABP | 147,116 | 53,043 |

Various sub-experiments were performed under Experiment 1 to determine the best U-Net architecture as an autoencoder (at least for this study). Their respective Mean Absolute Error (MAE) was recorded in each case.

**Variable Depth of the Encoder:** The depth (number of levels) of the U-Net was varied from 1 to 4 to determine whether the depth of the architecture had any effect on the extracted latent features from the autoencoder.

**Variable Width of the Encoder and Number of Features:** The width of the encoder, which represents the number of kernels or filters present in the input layer, was varied from 32 to 256.

**Variable Kernel Size:** The Kernel size was varied from 3 to 11 to see the effect of Kernel size on performance.

**Variable Number of Channels:** Four combinations of the four predictor signals were used for BP prediction. For one channel: only PPG, for two channels: PPG and ECG, for three channels: PPG and its two derivatives, and for four channels: all four types of signals were utilized.

**Experiments on Regression Techniques:** The extracted features were used to train some traditional Machine Learning regression techniques, namely, MLP, SGD, SVR, XgBoost, GradBoost, AdaBoost, K-Nearest Neighbor, and Random Forest to predict BP.

**BP Prediction from PPG-to-PPG Feature Mapping:** Apart from the primary approach of this study which aimed at mapping PPG and ECG features to ABP features for BP prediction, an additional experiment was performed aiming at predicting BP by mapping PPG (or PPG and ECG) to PPG i.e., PPG is taken as the target signal instead of ABP while using the same BP labels and ground truths. The significance of this study lies in taking the ABP signal completely out of the equation which would help avoid acquiring simultaneous ABP data during data acquisition and BP can be predicted from PPG alone.

B. Experiment 2 (Validating on External "BCG Dataset"):

The external BCG dataset has been investigated using two different methods. Firstly, the model trained on the whole UCI dataset is evaluated on the full BCG dataset (Method 1). Secondly, similar to Experiment 1 was performed on the UCI dataset, i.e., the model was trained using the BCG dataset through 5-Fold Cross-Validation (Method 2).

**Train on UCI, Test on BCG Dataset (Method 1):** In this experiment, the BCG Dataset was tested against a model trained on the whole UCI Dataset. The outcome from this experiment proved the performance and generalizability of a model trained using the proposed shallow U-Net-based autoencoder on a completely unknown dataset. The training and testing sets used for this experiment are described in Table 3.

**Table 3.** Description of Train and Test Sets for Experiment 2 (Method 1)

| No. of Channels | Channels | Target | Total Samples in the Train Set from UCI | Total Samples in the Test Set from BCG |
|---|---|---|---|---|
| 1 | PPG | ABP | 200,159 | 1,872 |
| 2 | PPG, ECG | ABP | 200,159 | 1,872 |
| 3 | PPG, VPG, APG | ABP | 200,159 | 1,872 |
| 4 | PPG, VPG, APG, ECG | ABP | 200,159 | 1,872 |

**5-Fold CV on BCG Dataset (Method 2):** The BCG dataset was divided into train-test fold (80:20) and validated using a 5-Fold Cross-Validation approach. The training set, in this case, contained 1,498 samples while the test set contained 374 samples (Table 4).

**Table 4.** Description of Train and Test Sets for Experiment 2 (Method 2)

| No. of Channels | Channels | Target | Total Samples in the Train Set | Total Samples in the Test Set |
|---|---|---|---|---|
| 1 | PPG | ABP | 1,498 | 374 |
| 2 | PPG, ECG | ABP | 1,498 | 374 |
| 3 | PPG, VPG, APG | ABP | 1,498 | 374 |
| 4 | PPG, VPG, APG, ECG | ABP | 1,498 | 374 |

## C. Evaluation Metrics

**Primary Evaluation Metric:**
Mean Absolute Error (MAE) was used as the primary evaluation metric for this study. For example, for predicted values $\hat{Y} = [\hat{y}_1, \hat{y}_2, \hat{y}_3, \dots, \hat{y}_n]$ and ground truth values $Y = [y_1, y_2, y_3, \dots, y_n]$, MAE is defined as [57]:

$$MAE = \frac{\sum_{i=1}^{n}|y_i - \hat{y}_i|}{n} \dots \dots \dots (1)$$

**British Hypertension Society (BHS) Standard:**
The British Hypertension Society (BHS) introduced a structured protocol [58] to act as a standard for assessing BP measuring devices and methods which has been frequently used in the literature as a metric. The BHS standard evaluates the performance based on absolute error while classifying the outcomes mainly into three categories viz. Grade A, B, and C. The grades are provided by measuring what percentage of the prediction absolute errors fall under (less than or equal to) 5 mmHg, 10 mmHg, and 15 mmHg, respectively. It is to be mentioned that for an algorithm or pipeline to obtain a certain grade, it has to satisfy the criteria of all three categories. There is also a Grade D for studies that fail to meet the requirements for Grade C [58].

**Association for the Advancement of Medical Instrumentation (AAMI) Standard:**
Association for the Advancement of Medical Instrumentation (AAMI) has proposed a similar standard [59] as BHS for evaluating BP measuring devices and algorithms. According to this standard, BP measuring systems should have a Mean Error (ME) and Standard Deviation (STD) of magnitude (absolute value) less than or equal to 5 mmHg and 8 mmHg, respectively. Moreover, the number of subjects to be evaluated should be greater than or equal to 85.

**Hypertension Classification:**
When the model is applied to a practical application, it will be quite useful to see if it can robustly classify the state of hypertension of the patient. This classification can be done from the values of SBP and DBP based on some simple scales or ranges provided in [60], which are the three most common cases viz. Normotension, Prehypertension, and Hypertension. Normotension ranges for DBP less than or equal to 80 and SBP less than or equal to 120. Prehypertension ranges for DBP between 80 and 90 inclusive and SBP between 120 and 140 inclusive, and Hypertension is ranged for DBP greater than 90 and SBP greater than 140. For Hypertension Classification, we used traditional evaluation metrics such as Precision, Recall, and F1-Score. If tp, tn, fp, and fn are used to denote true positives, true negatives, false positives, and false negatives, respectively, these metrics can be formulated as [61],

$$Precision = \frac{tp}{tp + fp} \quad \ldots \ldots \ldots (2)$$

$$Recall = \frac{tp}{tp + fn} \quad \ldots \ldots \ldots (3)$$

$$F1 - Score = \frac{2 \times tp}{2 \times tp + fp + fn} \quad \ldots \ldots \ldots (4)$$

**Statistical Analyses:**
Mainly two types of statistical analyses were performed in this study viz. Linear Regression and the Bland-Altman Plots [62]. The linear regression plots show the correlation between the ground truths and the predictions and can be represented by Equation 5 [63].

$$Y_i = \beta_0 + \beta_1 X_i \quad \ldots \ldots \ldots (5)$$

Here, $Y_i$ and $X_i$ are the dependent and independent variables, respectively. $\beta_0$ is the offset or the y-intercept and $\beta_1$ is the slope. The most positive correlation results in a slope of 1, which in turn varies between -1 and 1. In this study, we also represent the linear correlation performance with the Pearson Correlation Coefficient (PCC). PCC is the covariance of the two variables divided by the product of their standard deviations, as shown in Equation 6 [63].

$$r_{xy} = \frac{\sum_{i=1}^{n}(x_i - \bar{x})(y_i - \bar{y})}{\sqrt{\sum_{i=1}^{n}(x_i - \bar{x})^2} \sqrt{\sum_{i=1}^{n}(y_i - \bar{y})^2}} \quad \ldots \ldots \ldots (6)$$

Here, it is to be mentioned that the PCC formula for an entire population and a sample of the population is different due to considering population and sample means respectively during computation. In this case, PCC formulae for a sample have been used since the dataset is a sample of the originally collected dataset in MIMIC-II. On the contrary, we also computed and plotted the Bland-Altman plots to show the difference between the ground truths and the predictions over the whole BP range, which cannot be reflected upon properly from normal correlation plots.

## IV. Results
### A. Experiment 1: Train and Test on UCI Dataset
Several different studies were carried out in Experiment 1 as mentioned earlier to identify the best network architecture with optimized parameters. In what follows, we will report the results of these studies.

**Variable Depth of the Encoder:** As shown in Table 5, the MAE for BP prediction increased as the depth of the encoder increased. Based on this direct correlation, we can conclude that as the encoder became deeper, it more and more looked into complex features of the signals and the network became lesser efficient in capturing peripheral features such as SBP and DBP. For this reason, the shallowest version of U-Net as an autoencoder model performed best for BP prediction.

**Table 5.** MAE of BP Prediction for Variable Encoder Depth

| Fixed Parameters | Encoder Levels | MAE | |
| --- | --- | --- | --- |
| | | SBP | DBP |
| **Encoder Type:** U-Net<br>**Encoder Width:** 128 | 1 | 2.333 | 0.713 |

| Kernel Size: 3<br>No. of Channels: 4<br>No. of Extracted Feature: 1024<br>Regressor: MLP | 2 | 3.169 | 1.099 |
| --- | --- | --- | --- |
| | 3 | 3.763 | 1.243 |
| | 4 | 4.416 | 1.419 |

**Variable Width of the Encoder and Number of Features:** As shown in Figure 5, the width of the input layer of the encoder was varied from 32 to 256. The best performance was recorded at 128. The performance improved until 128 then started to drop again as the network becomes very wide and heavier than necessary. Here, the fixed parameters were Encoder Type, Encoder Depth, Kernel Size, Number of Channels, and Regressor Type.

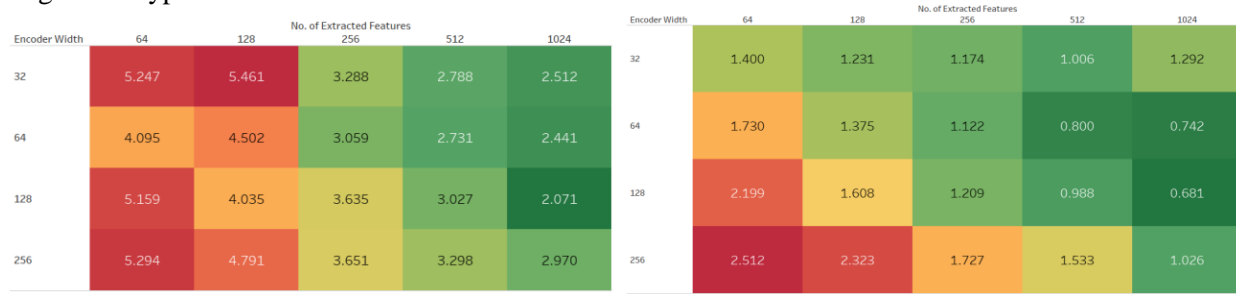

**Figure 5.** Heatmap Depicting MAE for SBP (left) and DBP (right) prediction while varying Encoder Width and Number of extracted features.

The U-Net-based autoencoder was used to extract features from both train and test sets. So, the optimal number of features to be extracted is also crucial to investigate. Figure 5 also reveals that the performance gets better until 1024 features, then start dropping. There can be a misconception that more features will provide better accuracy indefinitely. But in this case, it was noticed that the performance does not increase, rather drops a bit when the feature number is increased from 1024 to 2048 and the process becomes computationally expensive.

**Variable Number of Channels:** It can be noticed from Table 6 that performance improves by around 45% when two or three channels are used instead of using only PPG. The performance of the two and three-channel approaches are similar, while the performance improves again by around 25% when all four signals are used in combination. The same pattern was seen for both SBP and DBP even though SBP performed worse than DBP in all cases, which is a typical observation from the literature as well [12-16, 18, 20-27, 64].

**Table 6.** MAE of BP Prediction for Variable Channels

| Fixed Parameters | No. of Channels | MAE | |
| --- | --- | --- | --- |
| | | SBP | DBP |
| **Encoder Type:** U-Net<br>**Encoder Depth:** 1<br>**Encoder Width:** 128<br>**Kernel Size:** 3<br>**No. of Extracted Feature:** 1024<br>**Regressor:** MLP | 1 | 4.971 | 1.361 |
| | 2 | 2.513 | 0.825 |
| | 3 | 2.739 | 0.960 |
| | 4 | **2.333** | **0.713** |

One significant outcome from this experiment is that PPG and its first two derivatives perform similarly to PPG alone with ECG for BP prediction. So, ECG can be replaced just by deriving two derivatives of the

PPG signal and supplying them as two additional channels in U-Net. Removing ECG while maintaining the performance greatly reduces the complexity of the test setup.

**Variable Kernel Size:** The Kernel Size, k = 3 performed best as the Kernel size was varied from 3 to 11. The performance dropped as the Kernel Size was increased (Table 7).

**Table 7.** MAE of BP Prediction for Variable Kernel or Filter Size

| Fixed Parameters | Kernel Size | MAE | |
| --- | --- | --- | --- |
| | | SBP | DBP |
| **Encoder Type:** U-Net | 1 | 2.387 | 0.876 |
| **Encoder Depth:** 1 | 3 | **2.333** | **0.713** |
| **Encoder Width:** 128 | 5 | 2.503 | 0.949 |
| **No. of Channels:** 4 | 7 | 2.900 | 0.888 |
| **No. of Extracted Feature:** 1024 | 9 | 3.421 | 1.568 |
| **Regressor:** MLP | 11 | 4.544 | 1.388 |

Based on these experiments, the best U-Net architecture as an autoencoder is shown in Figure 6 along with annotations for all parameters. Here, in the bottom layer of the U-Net, an extra fully connected dense layer has been inserted to extract features. The number of parameters in the dense layer depends on the CNN block before it and the number of features to be extracted. For example, while extracting 1024 features, the size of the dense layer was (512*128*1024) = 67108864, which adds up to the size of the whole model. It is worth mentioning that the dense layer could be placed between CNN blocks of 512 by 256 which would double the number of parameters (512*256*1024), but doing it did not improve the performance.

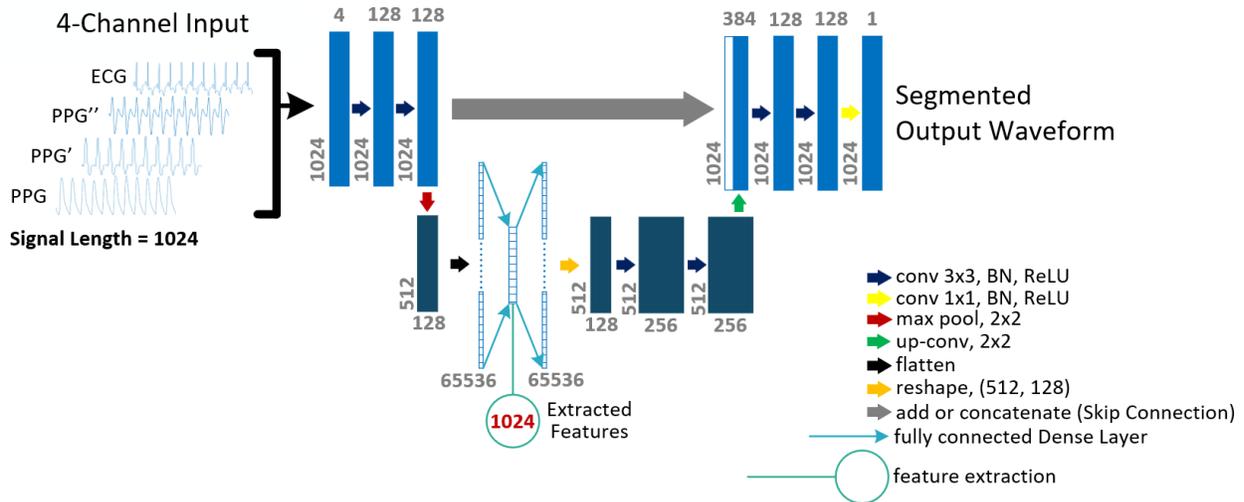

**Figure 6:** Architecture of the shallow U-Net model for feature extraction.

**Experiments on Regression Techniques:** The extracted features from the best autoencoder architecture, were trained using some traditional Machine Learning regression techniques to predict BP. As is evident from Table 8, Multi-Layer Perceptron (MLP) outperformed other classical machine learning techniques.

**Table 8.** Mean Absolute Error (MAE) of SBP and DBP for different ML techniques in Experiment 1

| Fixed Parameters | Regressor Algorithm | MAE for SBP | MAE for DBP |
| --- | --- | --- | --- |
| **Encoder Type:** U-Net | MLP | **2.333** | **0.713** |
| **Encoder Depth:** 1 | GradBoost | 5.837 | 1.418 |
| **Encoder Width:** 128 | SGD | 5.945 | 2.261 |
| **Kernel Size:** 3<br>**No. of Channels:** 4<br>**No. of Extracted Feature:** 1024 | SVM | 5.980 | 2.269 |
| | XgBoost | 6.089 | 1.429 |
| | K-Nearest Neighbor | 6.543 | 1.510 |
| | AdaBoost | 8.584 | 2.234 |

**BP Prediction from PPG-to-PPG Feature Mapping:** From Supplementary Table 1, it can be seen that the PPG-to-PPG approach to predict BP was not very successful, at least using this pipeline, due to lower correspondence between BP values and PPG patterns. This mini-experiment indirectly ascertained the robustness of the proposed pipeline in predicting BP by exploiting the relationship between BP values and corresponding ABP waveform patterns.

### BHS Standard

The criteria of the three grades along with the model performance of this study are presented in Table 9. From Table 9, it can be seen that with the developed pipeline, we have achieved Grade A for both SBP and DBP. Especially for DBP prediction, almost 100% of the signals met the Grade A criterion.

**Table 9.** Evaluation of BP Prediction in Experiment 1 in terms of BHS Standard

| | | Cumulative Error Percentage | | |
|---|---|---|---|---|
| | | ≤ 5 mmHg | ≤ 10 mmHg | ≤ 15 mmHg |
| **Our Results** | SBP | 92.02% | 99.18% | 99.85% |
| | DBP | 99.01% | 99.91% | 100.0% |
| **BHS Metric** | Grade A | 60% | 85% | 95% |
| | Grade B | 50% | 75% | 90% |
| | Grade C | 40% | 65% | 85% |

Histograms of MAE for SBP and DBP predictions for all of them are plotted in Supplementary Figure 2 (a). It can be seen that for the DBP, the MAE for almost all predictions is below or equal to 5 mmHg, which is the Grade A threshold. On the other hand, for SBP, MAE of most of the predictions is below or equal to 5 mmHg, which is BHS Grade A, and MAE of almost all predictions is below or equal to 10 mmHg, which is BHS Grade B.

### AAMI Standard:

As presented in Table 10, the predictions from our pipeline meet both categories of the AAMI standard keeping a large margin with the criteria.

**Table 10.** Evaluation of BP Prediction in Experiment 1 in terms of AAMI Standard

| | | ME (mmHg) | STD (mmHg) | Number of Subjects |
|---|---|---|---|---|
| **Our Results** | SBP | 0.09 | 0.94 | 942 |
| | DBP | -0.019 | 2.876 | |
| **AAMI Standard** | | ≤ 5 mmHg | ≤ 8 mmHg | ≥ 85 |

Error measurements for all SBP and DBP predictions are plotted in Supplementary Figure 2 (b). It can be seen that the error is normally distributed following the Central Limit Theorem. The SBP predictions are more widely distributed than the DBP predictions implying their higher deviation and lower accuracy.

### Hypertension Classification Performance:

Table 11 reports the corresponding SBP and DBP prediction performance for each of the hypertension classes in terms of Precision, Recall, and F1-score. From Table 11 and the corresponding Confusion Matrix shown in Figure 7, it can be seen that the model performs best for the subjects who are within the normotension range, i.e., without any hypertension. In terms of DBP, almost no misclassification is noticed among the samples belonging to this group. The worst performance is noticed in the pre-hypertension group. The prehypertension cases got misclassified as both normotension and hypertension cases, more frequently with the hypertension ones; this could be due to it being a group in the middle. Notably, some hypertension subjects got misclassified as pre-hypertension cases as well. Among 53,043 test signals, around 92% of the cases belong to the normotension group while for SBP, the distribution is more uniform which can also be visualized from Figure 7. The overall accuracy for the DBP prediction is about 98.95% whereas it is 94.14% for SBP.

**Table 11.** BP Classification Metrics for Various Hypertension Levels – Experiment 1

| Class | DBP | | | | SBP | | | |
|---|---|---|---|---|---|---|---|---|
| | Range | Precision | Recall | F1-score | Range | Precision | Recall | F1-score |
| **Normotension** | DBP ≤ 80 | 99.56% | 99.54% | 99.55% | SBP ≤ 120 | 95.59% | 94.69% | 95.14% |
| **Prehypertension** | 80 < DBP ≤ 90 | 90.47% | 91.88% | 91.17% | 120 < SBP ≤ 140 | 91.70% | 91.72% | 91.71% |
| **Hypertension** | 90 < DBP | 96.33% | 93.15% | 94.71% | 140 < SBP | 95.38% | 96.06% | 95.72% |

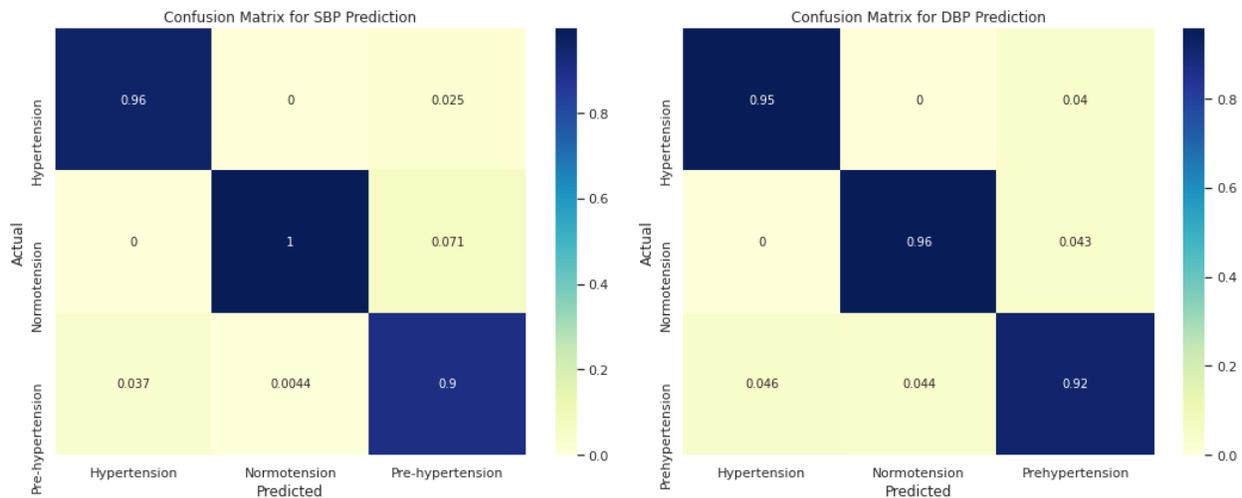

**Figure 7:** Confusion Matrix for Classifying Various Hypertension Levels based on SBP (left) and DBP (right) Predictions

**Statistical Analysis:**

The response plots for SBP and DBP regression outcomes are shown in Figure 8(a). From the plots, a high correlation between the target values and the ground truths is evident. The Pearson Correlation Coefficients for SBP and DBP predictions are 0.991 and 0.996, respectively, indicating a strong positive correlation between the target variables and the ground truths for both cases. On the other hand, p-values of approximately 0.01 for both cases indicate the statistical significance of the outcomes of this experiment when the test set contains 53,043 samples. Thus, the Null Hypothesis, which got rejected, stated that there is no relation between the predictions and the ground truths.

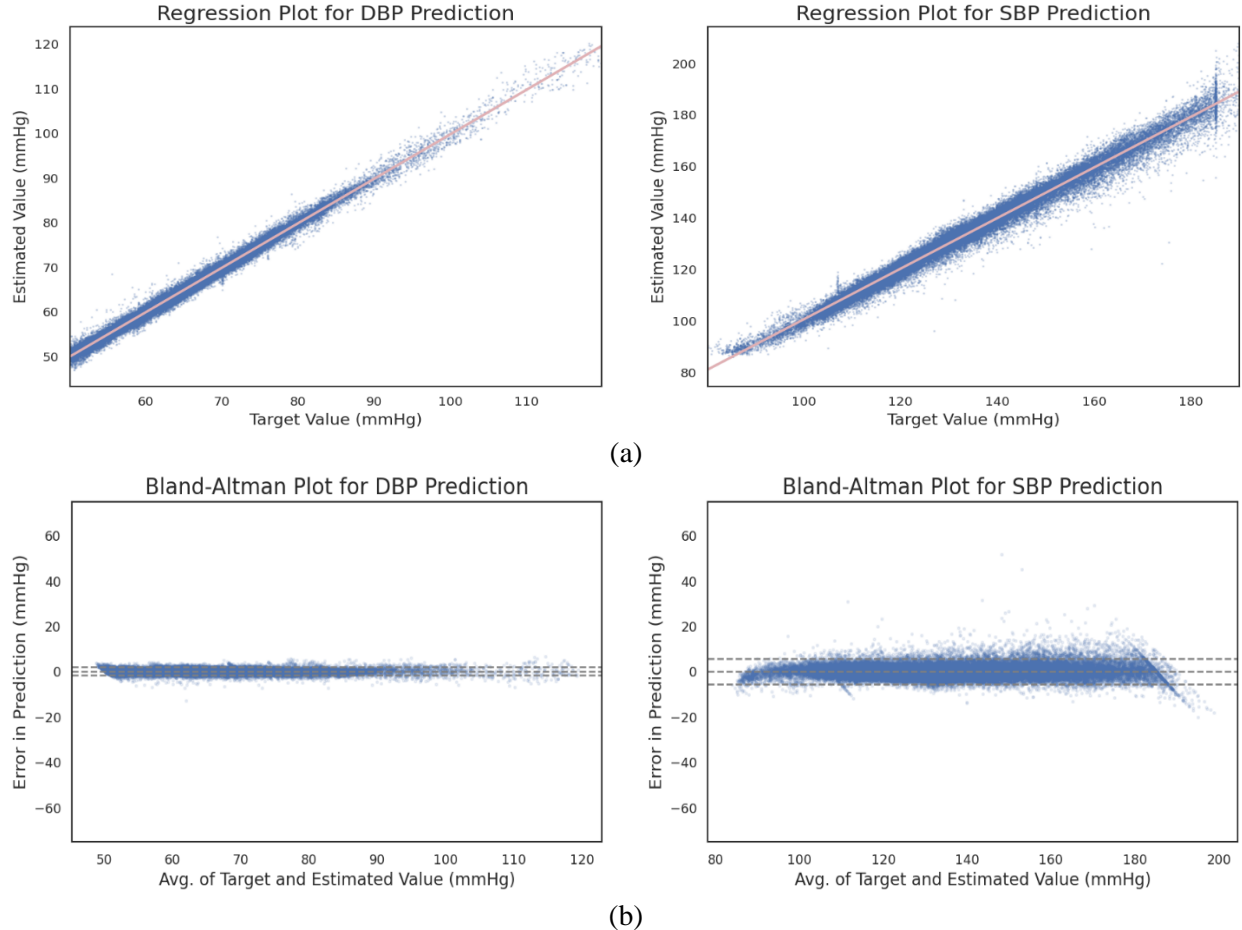

**Figure 8:** Regression Plots for DBP (left) and SBP (right) Predictions vs. Respective Ground Truths (a) and Bland-Altman Plots for DBP (left) and SBP (right) Predictions (b).

Figure 8(b) represents the Bland-Altman plots for DBP and SBP predictions, respectively. The 95% significance level, which is shown by the dashed lines, spans the segment from $\mu - 1.96\sigma$ to $\mu + 1.96\sigma$, where $\mu$ and $\sigma$ are population mean and standard deviation of the distribution, respectively. For SBP and DBP, the means are 5.618 and 1.933, respectively while the standard deviations are 2.89 and 0.894, respectively. So, SBP and DBP spanned within the range [-0.046:11.282] and [0.181:3.685], respectively. It can be understood from Figure 8 that even though SBP samples deviated more (which is expected), in both cases most error terms fell within the dash marked 5 mmHg range. The presence of outliers is not severe, in fact very low for DBPs. Another important observation from the Bland-Altman plot is that the error magnitudes remain almost similar over the SBP and DBP ranges. So, the error performances of ABP signals with extreme BP values (severe hypertension) were not affected by their high magnitude.

### B. Experiment 2 (Validating on an External "BCG" Dataset):
In this experiment, after training on the whole pre-processed UCI dataset, the created model has been tested on the whole (similarly pre-processed) BCG dataset. The main aim was to prove the effectiveness of the shallow U-Net model trained on a large dataset on an external dataset.

**Performance Evaluation:**
MAE for SBP and DBP was found to be **2.728** and **1.166,** respectively after testing the whole BCG dataset with 1,872 samples by the model trained on the whole UCI dataset. MAE was slightly higher than the results obtained from Experiment 1 with the UCI dataset but still better than any past study. The performance is excellent considering that the BCG dataset is completely unknown compared to MIMIC-II (UCI repository) from all aspects from data acquisition setup to data pre-processing. But when there was no transfer learning, the MAE for 5-Fold CV on the BCG dataset was found to be **6.336** and **2.658** for SBP and DBP, respectively. It can imply that the autoencoder requires a good amount of nicely varying balanced datasets to extract quality features. So, it is important to train the proposed model using a large, general dataset that contains an ample number of features. Note that BHS and AAMI metrics information for external validation have not been provided since the number of patients in the BCG dataset does not suffice the minimum requirements for these metrics.

### C. Comparison with Existing Works:

Various research groups around the world attempted to predict BP from PPG and ECG signals separately or in combination using various machine learning techniques. It is hard to directly compare and evaluate the performances of those studies due to multiple factors such as the number of patients, data pre-processing, signal length, machine learning models, so on and so forth. In Table 12, only papers reporting their error performance in Mean Absolute Error (MAE) have been reported. The entries in Table 12 are sorted in ascending order by the year of publication of the respective papers. Some works have low performance in terms of BHS and other metrics due to high standard deviation even though their error is low, which are reported in Table 13.

**Table 12.** Comparison of Past Studies based on Mean Absolute Error (MAE) Performance

| Study | Year Published | Dataset | Input Signals | Method | MAE (mmHg) | |
|---|---|---|---|---|---|---|
| | | | | | SBP | DBP |
| Kurylayak et al. [20] | 2013 | MIMIC, 15000 Pulsations | PPG | ANN | 3.80 | 2.21 |
| Kachuee et al. [12] | 2015 | MIMIC II (851 Subjects) | PPG, ECG | SVR | 12.38 | 6.34 |
| Kachuee et al. [14] | 2017 | MIMIC-II (942 Subjects) | PPG, ECG | Adaptive Boosting | 11.17 | 5.35 |
| Wang et al. [21] | 2018 | MIMIC, 72 Subjects | PPG | ANN | 4.02 | 2.27 |
| Xie et al. [15] | 2018 | University of Queensland Vital Signs Dataset [36] | PPG | Random Forest | 4.21 | 3.24 |
| Slapničar et al. [64] | 2018 | 22 Healthy Subjects [38] | PPG | Ensemble of Regression Trees | 6.70 | 4.42 |
| Slapničar et al. [26] | 2019 | MIMIC, 510 Subjects | | Spectro-Temporal ResNet | 9.43 | 6.88 |
| Manamperi et al. [22] | 2019 | MIMIC-II | PPG | ANN | 4.8 | 2.5 |
| | | 50 Healthy Subjects | | | 4.1 | 1.7 |
| Sasso et al. [16] | 2020 | HYPE (12 Subjects) | PPG, ECG | Gradient Boosting | 8.79 | 6.37 |
| | | EVAL (26 Subjects) [17] | | | 14.74 | 7.12 |
| Esmaelpoor et al. [38] | 2020 | MIMIC-II (200 Subjects) | PPG | CNN + LSTM | 1.91 | 0.67 |
| Ibtehaz et al. [13] | 2020 | MIMIC-II (942 Subjects) | PPG | CNN + CNN | 5.73 | 3.45 |

| Chowdhury et al. [18] | 2020 | 222 Recordings from 126 Subjects [19] | PPG | Gaussian Process Regression (GPR) | 3.02 | 1.74 |
|---|---|---|---|---|---|---|
| Li et al. [24] | 2020 | MIMIC-II (3000 Recordings from UCI Repository) | PPG, ECG | LSTM | 4.63 | 3.15 |
| Hsu et al. [23] | 2020 | MIMIC-II (9000 Recordings from UCI Repository) | PPG, ECG | ANN | 3.21 | 2.23 |
| Athaya et al. [27] | 2021 | MIMIC-II (100 Subjects) | PPG | CNN | 3.68 | 1.97 |
| Harfiya et al. [25] | 2021 | MIMIC-II (5289 Recordings from UCI Repository) | PPG | LSTM | 4.05 | 2.41 |
| This Study | 2021 | MIMIC-II (942 Subjects – 12000 Recordings from UCI Repository) | PPG, ECG | CNN + ANN | **2.333** | **0.713** |
| | | MIMIC-II + BCG (942+40 = 982 Subjects) | | | **2.728** | **1.166** |
| **AAMI Standard** | | | | | | ≤ 5 |

Note: It is important to mention that Hsu et al. [23] in their paper reported that they used 9,000 subjects' data for BP prediction from the UCI repository, but it was 9,000 out of 12,000 instances or recordings of data collected from the MIMIC-II dataset. It is the data from 942 patients as reported by Kauchee et al. [14], the originator of this dataset. A similar case happened for the case of Harfiya et al. [25] where they reported 5,289 signal instances from the UCI repository as 5,289 patients. In comparison, this study fully utilized all 12,000 instances.

Performance metrics such as MAE do not always show the complete picture of the performance of a study. For this reason, many studies in this domain represent their results in terms of BHS metrics. A comparison of BHS metrics of the current work with some past studies has been shown in Table 13. As it can be seen, only a handful of very recent studies could reach BHS Grade A for both SBP and DBP predictions. It is noticeable from Tables 12 and 13 that even though some recent studies gained close or even better MAE than this study, they have lower performance in BHS metrics due to high deviation in the result (can be further confirmed by comparing the respective AAMI metric). In terms of BHS, AAMI, and other metrics, our performance is the best so far, even with a larger dataset than almost all of these studies in terms of total signal duration. Moreover, the best performing shallow U-Net architecture proposed in this study as an autoencoder is also very lightweight. For example, the level-4, general version of U-Net used by Ibtehaz et al. [13] has approximately 10.5 million parameters without Deep Supervision (and it is just one of the two CNN networks used in the pipeline, the other one being the MultiResUNet [29]) while the shallow, level-1 U-Net model used for this experiment has only around 0.55M parameters, around 19 times lighter.

**Table 13.** Comparison of Past Studies based on their Performance of BHS Metrics

| Study | SBP (%) in BHS Metrics | | | | DBP (%) in BHS Metrics | | | |
|---|---|---|---|---|---|---|---|---|
| | Grade A | Grade B | Grade C | Attained Grade | Grade A | Grade B | Grade C | Attained Grade |
| Kachuee et al. [12] | 29 | 52 | 70 | D | 51 | 79 | 94 | B |
| Kachuee et al. [13] | 34 | 57 | 73 | D | 63 | 87 | 96 | A |

| Xie et al. [15] | 74 | 91 | 95 | A | 80 | 94 | 98 | A |
| Mousavi et al. [65] | 71 | 77 | 84 | D | 84 | 92 | 97 | A |
| Esmaelpoor et al. [38] | 74 | 94 | 98 | A | 93 | 99 | 100 | A |
| Ibtehaz et al. [13] | 71 | 85 | 91 | B | 83 | 92 | 96 | A |
| Li et al. [24] | 60 | 80 | 89 | B | 77 | 96 | 100 | A |
| Hsu et al. [23] | 81 | 96 | 98 | A | 90 | 98 | 100 | A |
| Athaya et al. [27] | 76 | 94 | 99 | A | 94 | 99 | 100 | A |
| Harfiya et al. [25] | 71 | 94 | 99 | A | 91 | 99 | 100 | A |
| **This Study** | **92** | **99** | **99** | **A** | **99** | **~100** | **~100** | **A** |

*%: The percentage of predicted signals falling within 5 (Grade A), 10 (Grade B), and 15 (Grade C) mmHg of their respective ground truth signals, respectively.

## V. Conclusion

This study aimed at developing a novel pipeline for BP prediction from PPG and ECG signals by experimenting with the U-Net architecture being used as an autoencoder to extract optimum features. Instead of the raw signals, the extracted features were regressed using machine learning techniques to predict SBP and DBP. The strength of this work lies in how the U-Net architecture was utilized for feature extraction, thereby achieving the best performance from the shallowest version of the U-Net architecture on the current largest possible dataset from the UCI repository. The extracted features were so efficient in predicting the SBP and DBP that there is a significant performance boost compared to any previous study. Our lightweight network can be helpful for deployment in a resource-constrained setting. Independent test sets were used for evaluation purposes for both experiments performed in this study proving the robustness of the proposed pipeline. The dataset used for the second experiment was a dataset acquired through a completely different process (e.g., ABP was recorded non-invasively), but still our model achieved high performance when evaluated thereon thereby showing the generalizability thereof. This strongly suggests that extracting features from this large dataset using the shallow autoencoder provided the trained model with enough generalizable features to perform robustly even on external datasets. Some studies (e.g., [13]) reported that avoiding ECG signals as the second predictor, while maintaining high performance, could help in simplifying the hardware design, device implementation, and patient monitoring. The current study showed that even without the ECG signal, the model can perform similarly by just using the first two derivatives of PPG instead. MAE for SBP and DBP predictions with three channels were 2.74 and 0.96, respectively, which is still one of the best performances so far compared to the past studies. Therefore, a three-channel model (PPG and two derivatives) can easily be used for deployment without any ECG signal provided that the model is trained on a large general dataset (like the UCI dataset). As the current approach did not design to handle motion artifacts in the acquired signals, it would be challenging to direct use the current system for a wearable device. Since the model was mostly trained on very clean signals, it could highly affect the model performance. However, the motion artifact can be corrected through many ways, one of which is proposed for the PPG signals of our research team [66]. In conclusion, the proposed model and framework can be suitable for deployment in remote monitoring servers and mobile applications for real-time non-invasive BP monitoring applications as the proposed model has only 0.5 million trainable parameters.

## Acknowledgment
This work was supported in part by the Qatar National Research Fund under Grant NPRP12S-0227-190164 and in part by the International Research Collaboration Co-Fund (IRCC) through Qatar University under Grant IRCC-2021-001. The statements made herein are solely the responsibility of the authors. Open access publication is supported by Qatar National Library.

## Data Availability

The data used in this experiment along with other relevant documents used to complete this work have been provided in the following GitHub repository [67].

## References:

[1] "The top 10 causes of death", Who. int, 2021. [Online]. Available: https://www.who.int/news-room/fact-sheets/detail/the-top-10-causes-of-death.

[2] Heart Disease and Stroke. Cdc.gov. (2021). Retrieved 18 August 2021, from https://www.cdc.gov/chronicdisease/resources/publications/factsheets/heart-disease-stroke.htm.

[3] Bhatt, S., & Dransfield, M. (2013). Chronic obstructive pulmonary disease and cardiovascular disease. Translational Research, 162(4), 237-251. https://doi.org/10.1016/j.trsl.2013.05.001.

[4] Heart–Lung Interaction via Infection | Annals of the American Thoracic Society. Dx.doi.org. (2021). Retrieved 18 August 2021, from https://dx.doi.org/10.1513%2FAnnalsATS.201306-157MG.

[5] C. Wu, H. Hu, Y. Chou, N. Huang, Y. Chou, and C. Li, "High Blood Pressure and All-Cause and Cardiovascular Disease Mortalities in Community-Dwelling Older Adults", Medicine, vol. 94, no. 47, p. e2160, 2015. Available: 10.1097/md.0000000000002160.

[6] "Vital signs: awareness and treatment of uncontrolled hypertension among adults--the United States, 2003-2010", PubMed, 2021. [Online]. Available: https://pubmed.ncbi.nlm.nih.gov/22951452/.

[7] W. Organization, "A global brief on hypertension: silent killer, global public health crisis: World Health Day 2013", Apps.who.int, 2021. [Online]. Available: https://apps.who.int/iris/handle/10665/79059. [Accessed: 22- May- 2021].

[8] C. Goodman and G. Kitchen, "Measuring arterial blood pressure", Anaesthesia & Intensive Care Medicine, vol. 22, no. 1, pp. 49-53, 2021. Available: 10.1016/j.mpaic.2020.11.007.

[9] A. Meidert and B. Saugel, "Techniques for Non-Invasive Monitoring of Arterial Blood Pressure", 2021.

[10] K. Lakhal, S. Ehrmann and T. Boulain, "Noninvasive BP Monitoring in the Critically Ill", Chest, vol. 153, no. 4, pp. 1023-1039, 2018. Available: 10.1016/j.chest.2017.10.030.

[11] P. Salvi, A. Grillo, and G. Parati, "Noninvasive estimation of central blood pressure and analysis of pulse waves by applanation tonometry", Hypertension Research, vol. 38, no. 10, pp. 646-648, 2015. Available: 10.1038/hr.2015.78.

[12] Kachuee, M., Kiani, M., Mohammadzade, H., and Shabany, M., 2015. Cuff-less high-accuracy calibration-free blood pressure estimation using pulse transit time. 2015 IEEE International Symposium on Circuits and Systems (ISCAS).

[13] Ibtehaz, N.; Rahman, M.S. PPG2ABP: Translating Photoplethysmogram (PPG) Signals to Arterial Blood Pressure (ABP) Waveforms using Fully Convolutional Neural Networks. arXiv 2020, arXiv:2005.01669.

[14] M. Kachuee, M. Kiani, H. Mohammadzade and M. Shabany, "Cuffless Blood Pressure Estimation Algorithms for Continuous Health-Care Monitoring", IEEE Transactions on Biomedical Engineering, vol. 64, no. 4, pp. 859-869, 2017. Available: 10.1109/tbme.2016.2580904.

[15] Q. Xie, G. Wang, Z. Peng, and Y. Lian, "Machine Learning Methods for Real-Time Blood Pressure Measurement Based on Photoplethysmography", 2018 IEEE 23rd International Conference on Digital Signal Processing (DSP), 2018. Available: 10.1109/icdsp.2018.8631690.

[16] A. Morassi Sasso et al., HYPE: Predicting Blood Pressure from Photoplethysmograms in a Hypertensive Population BT - Artificial Intelligence in Medicine. Cham: Springer International Publishing, 2020.

[17] A. Esmaili, M. Kachuee, and M. Shabany, "Nonlinear cuff-less blood pressure estimation of healthy subjects using pulse transit time and arrival time," arXiv, pp. 1–10, 2018.

[18] M. H. Chowdhury et al., "Estimating blood pressure from the photoplethysmogram signal and demographic features using machine learning techniques," Sensors (Switzerland), vol. 20, no. 11, 2020, doi: 10.3390/s20113127.

[19]    Liang, G.L.Y.; Chen, Z.; Elgendi, M. PPG-BP Database. 2018. Available online: https://figshare.com/articles/PPG-BP_Database_zip/5459299/.
[20]    Y. Kurylyak, F. Lamonaca, and D. Grimaldi, "A Neural Network-based method for continuous blood pressure estimation from a PPG signal," Conf. Rec. - IEEE Instrum. Meas. Technol. Conf., pp. 280–283, 2013, doi: 10.1109/I2MTC.2013.6555424.
[21]    L. Wang, W. Zhou, Y. Xing, and X. Zhou, "A novel neural network model for blood pressure estimation using photoplethysmography without electrocardiogram," J. Healthc. Eng., vol. 2018, 2018, doi: 10.1155/2018/7804243.
[22]    B. Manamperi and C. Chitraranjan, "A robust neural network-based method to estimate arterial blood pressure using photoplethysmography.," Proc. - 2019 IEEE 19th Int. Conf. Bioinforma. Bioeng. BIBE 2019, pp. 681–685, 2019, doi: 10.1109/BIBE.2019.00128.
[23]    Y. C. Hsu, Y. H. Li, C. C. Chang, and L. N. Harfiya, "Generalized deep neural network model for cuffless blood pressure estimation with photoplethysmogram signal only," Sensors (Switzerland), vol. 20, no. 19, pp. 1–18, 2020, doi: 10.3390/s20195668.
[24]    Y. H. Li, L. N. Harfiya, K. Purwandari, and Y. Der Lin, "Real-time cuffless continuous blood pressure estimation using deep learning model," Sensors (Switzerland), vol. 20, no. 19, pp. 1–19, 2020, doi: 10.3390/s20195606.
[25]    L. N. Harfiya, C. C. Chang, and Y. H. Li, "Continuous blood pressure estimation using exclusively photoplethysmography by lstm-based signal-to-signal translation," Sensors, vol. 21, no. 9, 2021, doi: 10.3390/s21092952.
[26]    G. Slapni Č Ar, N. Mlakar, and M. Luštrek, "Blood pressure estimation from photoplethysmogram using a spectro-temporal deep neural network," Sensors (Switzerland), vol. 19, no. 15, 2019, doi: 10.3390/s19153420.
[27]    T. Athaya and S. Choi, "An estimation method of continuous non-invasive arterial blood pressure waveform using photoplethysmography: A u-net architecture-based approach," Sensors, vol. 21, no. 5, pp. 1–18, 2021, doi: 10.3390/s21051867.
[28]    "U-Net: Convolutional Networks for Biomedical Image Segmentation", Lmb.informatik.uni-freiburg.de, 2021. [Online]. Available: https://lmb.informatik.uni-freiburg.de/people/ronneber/u-net/.
[29]    N. Ibtehaz and M. Rahman, "MultiResUNet: Rethinking the U-Net architecture for multimodal biomedical image segmentation", Neural Networks, vol. 121, pp. 74-87, 2020. Available: 10.1016/j.neunet.2019.08.025.
[30]    Z. Zhou, M. M. Rahman Siddiquee, N. Tajbakhsh, and J. Liang, "UNet++: A Nested U-Net Architecture for Medical Image Segmentation BT  - Deep Learning in Medical Image Analysis and Multimodal Learning for Clinical Decision Support," 2018, pp. 3–11.
[31]    P. Esser and E. Sutter, "A Variational U-Net for Conditional Appearance and Shape Generation Heidelberg Collaboratory for Image Processing," Proc. IEEE Comput. Soc. Conf. Comput. Vis. Pattern Recognit., pp. 8857–8866, 2018.
[32]    Z. Zhang, Q. Liu, and Y. Wang, "Road Extraction by Deep Residual U-Net", IEEE Geoscience and Remote Sensing Letters, vol. 15, no. 5, pp. 749-753, 2018. Available: 10.1109/lgrs.2018.2802944.
[33]    F. Isensee et al., "nnU-Net: Self-adapting framework for unet-based medical image segmentation," arXiv, 2018.
[34]    V. Iglovikov and A. Shvets, "TernausNet: U-Net with VGG11 encoder pre-trained on imagenet for image segmentation," arXiv, 2018.
[35]    D. Stoller, S. Ewert, and S. Dixon, "Wave-U-Net: A multi-scale neural network for end-to-end audio source separation," Proc. 19th Int. Soc. Music Inf. Retr. Conf. ISMIR 2018, pp. 334–340, 2018, doi: 10.5281/zenodo.1492417.
[36]    Ö. Çiçek, A. Abdulkadir, S. S. Lienkamp, T. Brox, and O. Ronneberger, "3D U-net: Learning dense volumetric segmentation from sparse annotation," Lect. Notes Comput. Sci. (including Subser. Lect. Notes Artif. Intell. Lect. Notes Bioinformatics), vol. 9901 LNCS, pp. 424–432, 2016, doi: 10.1007/978-3-319-46723-8_49.

[37] X. Hao, X. Su, Z. Wang, H. Zhang, and Batushiren, "Unetgan: A robust speech enhancement approach in the time domain for extremely low signal-to-noise ratio condition," Proc. Annu. Conf. Int. Speech Commun. Assoc. INTERSPEECH, vol. 2019-September, pp. 1786–1790, 2019, doi: 10.21437/Interspeech.2019-1567.
[38] J. H. Kim and J. H. Chang, "Attention Wave-U-Net for acoustic echo cancellation," Proc. Annu. Conf. Int. Speech Commun. Assoc. INTERSPEECH, vol. 2020-October, pp. 3969–3973, 2020, doi: 10.21437/Interspeech.2020-3200.
[39] X. Wu, M. Li, X. Lin, J. Wu, Y. Xi, and X. Jin, "Shallow triple Unet for shadow detection", Twelfth International Conference on Digital Image Processing (ICDIP 2020), 2020. Available: 10.1117/12.2572916.
[40] J. Esmaelpoor, M. H. Moradi, and A. Kadkhodamohammadi, "A multistage deep neural network model for blood pressure estimation using photoplethysmogram signals," Comput. Biol. Med., vol. 120, no. 350, p. 103719, 2020, doi: 10.1016/j.compbiomed.2020.103719.
[41] Dheeru, D. and Casey, G., 2017. UCI Machine Learning Repository. [online] Archive.ics.uci.edu. Available at: <http://archive.ics.uci.edu/ml>.
[42] Archive.physionet.org. 2021. MIMIC-II Databases. [online] Available at: <https://archive.physionet.org/mimic2/>.
[43] Physionet.org. 2021. MIMIC-III Waveform Database v1.0. [online] Available at: <https://physionet.org/content/mimic3wdb/1.0/>.
[44] C. Carlson, V. Turpin, A. Suliman, C. Ade, S. Warren, and D. Thompson, "Bed-Based Ballistocardiography: Dataset and Ability to Track Cardiovascular Parameters", Sensors, vol. 21, no. 1, p. 156, 2020. Available: 10.3390/s21010156.
[45] "NI-9220", Ni.com, 2021. [Online]. Available: https://www.ni.com/en-lb/support/model.ni-9220.html.
[46] "Finapres Medical Systems | Products - Finometer PRO", Finapres.com, 2021. [Online]. Available: https://www.finapres.com/Products/Finometer-PRO.
[47] Mathworks.com. 2021. Moving minimum - MATLAB movmin. [online] Available at: <https://www.mathworks.com/help/matlab/ref/movmin.html>.
[48] Mathworks.com. 2021. Polynomial curve fitting - MATLAB polyfit. [online] Available at: <https://www.mathworks.com/help/matlab/ref/polyfit.html>.
[49] Mathworks.com. 2021. Polynomial evaluation - MATLAB polyval. [online] Available at: <https://www.mathworks.com/help/matlab/ref/polyval.html>.
[50] Mohebbian, M., Dinh, A., Wahid, K. and Alam, M., 2020. Blind, Cuff-less, Calibration-Free, and Continuous Blood Pressure Estimation using Optimized Inductive Group Method of Data Handling. Biomedical Signal Processing and Control, 57, p.101682.
[51] Chakraborty, A., Sadhukhan, D., and Mitra, M., 2019. An Automated Algorithm to Extract Time Plane Features from the PPG Signal and its Derivatives for Personal Health Monitoring Application. IETE Journal of Research, pp.1-13.
[52] Elgendi, M., Liang, Y. and Ward, R., 2018. Toward Generating More Diagnostic Features from Photoplethysmogram Waveforms. Diseases, 6(1), p.20.
[53] Mathworks.com. 2021. Differences and approximate derivatives - MATLAB diff. [online] Available at: <https://www.mathworks.com/help/matlab/ref/diff.html>.
[54] Mathworks.com. 2021. Take Derivatives of a Signal- MATLAB & Simulink. [online] Available at: <https://www.mathworks.com/help/signal/ug/take-derivatives-of-a-signal.html>.
[55] Mathworks.com. 2021. Design digital filters - MATLAB designfilt. [online] Available at: <https://www.mathworks.com/help/signal/ref/designfilt.html>.
[56] Mathworks.com. 2021. Average filter delay (group delay) - MATLAB grpdelay. [online] Available at: <https://www.mathworks.com/help/signal/ref/grpdelay.html#f7-916897_sep_shared-n>.
[57] "Mean Absolute Error (MAE) ~ Sample Calculation", Medium, 2021. [Online]. Available: https://medium.com/@ewuramaminka/mean-absolute-error-mae-sample-calculation-6eed6743838a.

[58]     E. O'Brien, J. Petrie, W. Littler, M. de Swiet, P. L. Padfield, D. Altman, M. Bland, A. Coats, N. Atkins, et al., "The British hypertension society protocol for the evaluation of blood pressure measuring devices," J hyper tens, vol. 11, no. Suppl 2, pp. S43–S62, 1993.

[59]     A. for the Advancement of Medical Instrumentation et al., "American national standard. manual, electronic or automated sphygmomanometers ANSI," AAMI SP10-2002, Tech. Rep., 2003.

[60]     S. W. Holm, L. L. Cunningham, E. Bensadoun, and M. J. Madsen, "Hypertension: classification, pathophysiology, and management during outpatient sedation and local anesthesia," Journal of oral and maxillofacial surgery, vol. 64, no. 1, pp. 111–121, 2006.

[61]     K. Ping Shung, "Accuracy, Precision, Recall or F1?", Medium, 2021. [Online]. Available: https://towardsdatascience.com/accuracy-precision-recall-or-f1-331fb37c5cb9.

[62]     D. Giavarina, "Understanding Bland Altman analysis," Biochemia Medica: Biochemia Medica, vol. 25, no. 2, pp. 141–151, 2015.

[63]     "Simple Linear Regression and Pearson Correlation - StatsDirect", Statsdirect.com, 2021. [Online]. Available: https://www.statsdirect.com/help/regression_and_correlation/simple_linear.htm.

[64]     G. Slapničar and M. Luštrek, "Blood pressure estimation with a wristband optical sensor," UbiComp/ISWC 2018 - Adjun. Proc. 2018 ACM Int. Jt. Conf. Pervasive Ubiquitous Comput. Proc. 2018 ACM Int. Symp. Wearable Comput., pp. 758–761, 2018, doi: 10.1145/3267305.3267708.

[65]     S. S. Mousavi, M. Firouzmand, M. Charmi, M. Hemmati, M. Moghadam, and Y. Ghorbani, "Blood pressure estimation from appropriate and inappropriate PPG signals using A whole-based method," Biomed. Signal Process. Control, vol. 47, pp. 196–206, 2019, doi: 10.1016/j.bspc.2018.08.022.

[66]     Shuzan, Md Nazmul Islam, Moajjem Hossain Chowdhury, Md Shafayet Hossain, Muhammad EH Chowdhury, Mamun Bin Ibne Reaz, Mohammad Monir Uddin, Amith Khandakar, Zaid Bin Mahbub, and Sawal Hamid Md Ali. "A Novel Non-Invasive Estimation of Respiration Rate From Motion Corrupted Photoplethysmograph Signal Using Machine Learning Model." IEEE Access 9 (2021): 96775-96790.

[67]     "GitHub - Sakib1263/Shallow-UNet-based-Autoencoder-for-BP-Prediction: This repository contains files related to the paper called "A Shallow U-Net Architecture for Reliably Predicting Blood Pressure (BP) from Photoplethysmogram (PPG) and Electrocardiogram (ECG) Signals"", GitHub, 2021. [Online]. Available: https://github.com/Sakib1263/Shallow-UNet-based-Autoencoder-for-BP-Prediction. [Accessed: 21- Jul- 2021].